\documentstyle[epsfig,12pt]{article}
\def \bea{\begin{eqnarray}}
\def \beq{\begin{equation}}

\def \eea{\end{eqnarray}}
\def \eeq{\end{equation}}

\def \ko{K^0}

\def \ok{\overline{K}^0}

\def \pr{\parallel}

\begin{document}
\begin{titlepage}

\large
\centerline {\bf Comments on CKM Elements
\footnote{ Enrico Fermi Institute preprint EFI 2000-42, hep-ph/0011184.
To be published in Proceedings of Beauty 2000, Kibbutz Maagan, Israel,
September 13--18, 2000, edited by S. Erhan, Y. Rozen, and P. E. Schlein,
Nucl.\ Inst.\ Meth. A, 2001.}}
\normalsize
 
\vskip 2.0cm
\centerline {Jonathan L. Rosner~\footnote{rosner@hep.uchicago.edu}}
\centerline {\it Enrico Fermi Institute and Department of Physics}
\centerline{\it University of Chicago, 5640 S. Ellis Avenue, Chicago, IL 60637}
\vskip 4.0cm
 
\centerline {\bf Abstract}
\vskip 1.0cm

The sensitivity of determination of elements of the Cabibbo-Kobayashi-Maskawa
(CKM) matrix to experimental inputs (particularly $|V_{cb}|$) is discussed
and caution in assigning probable errors is urged.
\bigskip

\noindent
PACS Categories:  12.15.Hh, 12.15.Ff, 13.25.Hw, 14.65.Fy

\vfill
\end{titlepage}

\newpage

While experiments on magnitudes and phases of Cabibbo-Kobayashi-Maskawa
(CKM) matrix elements have become increasingly precise,
significant uncertainties remain in relating hadron information
to fundamental quark couplings.  We stress the effect of some of these
uncertainties, estimate the allowed parameter space, and conclude that it is
consistent with recent measurements of the CP asymmetry in $B \to J/\psi K_S$
\cite{BaO,BEO}.

We parametrize the CKM matrix \cite{WP} in terms of $\lambda \equiv V_{us}
\simeq 0.22$, $A \equiv |V_{cb}|/\lambda^2$, $\rho \equiv {\rm Re}(V_{ub})/
(A \lambda^3)$, and $\eta = - {\rm Im}(V_{ub})/(A \lambda^3)$.  The elements
$V_{ud}$, $V_{us}$, $V_{cd}$, and $V_{cs}$ are well-determined
\cite{FJG,GKR}.  Our concern will lie primarily with $V_{cb}$, $V_{ub}$,
and $V_{td}$.  Unitarity predicts $V_{ts} \simeq - V_{cb}$ and $V_{tb}
\simeq 1$.

Determination of $V_{cb}$ using inclusive semileptonic decays requires an
understanding of kinematic effects associated with $m_b$ and $m_c$.
The difference $m_b - m_c$ is constrained by spectroscopy to be close to
3.34 GeV \cite{JRTASI}, but a residual dependence on $m_b$ exists.
A free-quark estimate yields $|V_{cb}| = 0.0384 - 0.0008[(m_b - 4.7~{\rm GeV})
/(0.1 ~{\rm GeV})]$, for $\tau_b = 1.6$ ps and a branching ratio ${\cal B}
(B \to X_c \ell \nu) = 10.2\%$. Thus if $m_b$ is uncertain by 0.3 GeV (my
guess), $|V_{cb}|$ is uncertain by $\pm 0.0024$.

Determination of $V_{cb}$ using exclusive semileptonic decays (see, e.g,
\cite{CLEOVcb}) introduces a strong correlation between its value and that
of the form factor slope parameter $\rho^2$, so that values between 0.038
and 0.044 seem credible.  This is the range we shall adopt.  It is considerably
broader than that which appears in several reviews and in two presentations
\cite{FS} at this Conference.

The determination of $|V_{ub}|$ or $|V_{ub}/V_{cb}|$ relies upon semileptonic
charmless $B$ decays either giving rise to leptons beyond the charm limit
(and hence a tiny fraction of the spectrum), or a very precise modeling of
the leptons due to semileptonic $b \to c \ell \nu$ decay.  Values of
$|V_{ub}|$ ranging from 0.003 to 0.0045 have appeared in recent analyses.
In accord with \cite{GKR} and \cite{Flg}, we therefore take $|V_{ub}/V_{cb}|
= 0.09 \pm 0.025$ and hence $(\rho^2 + \eta^2)^{1/2} = 0.41 \pm 0.11$.

Loop diagrams describing CP-violating $\ko$--$\ok$ mixing and the mixing of
neutral nonstrange and strange $B$ mesons provide constraints on the elements
$V_{td}$ and $V_{ts}$.  We learn Im($V_{td}^2$), $|V_{td}|^2$, and $|V_{ts}/
V_{td}|$ from $\epsilon_K$, $\Delta m_d$, and $\Delta m_s/\Delta m_d$,
respectively.

Using standard expressions for loop diagrams \cite{IL} and QCD corrections
\cite{B97}, the parameter $\epsilon_K$ leads to the constraint
\beq
\eta \left[ 1 - \rho + B \left( \frac{m_c}{1.4~{\rm GeV}} \right)^2 \right]
= C \left( \frac{0.8}{B_K} \right)~~~,
\eeq
where $B=(0.46,0.39,0.28)$ and $C=(0.51,0.38,0.28)$ for $|V_{cb}| = (0.038,
0.041,$ 0.044).  A recent estimate of the vacuum-saturation factor $B_K$
\cite{Lubicz} gives $0.87 \pm 0.13$.  The value of $C$ changes by
nearly a factor of 2 over the range we consider possible for $|V_{cb}|$.
We plot the region of $(\rho,\eta)$ allowed by the simpler constraint
$\eta (1 - \rho + 0.39) = 0.35 \pm 0.12$.

The value of $\Delta m_d$, whose present world average is $\Delta m_d =
0.487 \pm 0.014$ ps$^{-1}$ \cite{mix}, is proportional to $f_B^2 B_B
|V_{td}|^2$, where $f_B$ is the $B$ meson decay constant and $B_B$ is the
vacuum saturation factor.  For $f_B \sqrt{B_{B}} = 230 \pm 40$ MeV \cite{lat}
and $0.038 \le |V_{cb}| \le 0.044$, one then finds $0.66 \le
|1 - \rho - i \eta| \le 1.08$.

The lower limit $\Delta m_s > 15$ ps$^{-1}$ may be interpreted in terms
of a constraint on $|V_{ts}/V_{td}|$ if the effects of flavor-SU(3) breaking
in matrix elements can be estimated.  Lattice estimates of the parameter
$\xi \equiv f_{B_s}\sqrt{B_{B_s}}/f_B\sqrt{B_B}= 1.1 \div 1.2$ are too
restrictive in my opinion; a quark
model obtains 1.25 for this ratio \cite{JRTASI,JRFM}, about $2 \sigma$ above
the lattice range.  With 1.25
as an upper limit one obtains the bound $|1-\rho-i\eta| < 1.01$.

\begin{figure}
% \vspace{2.8in}
\centerline{\epsfysize = 2.8in \epsffile {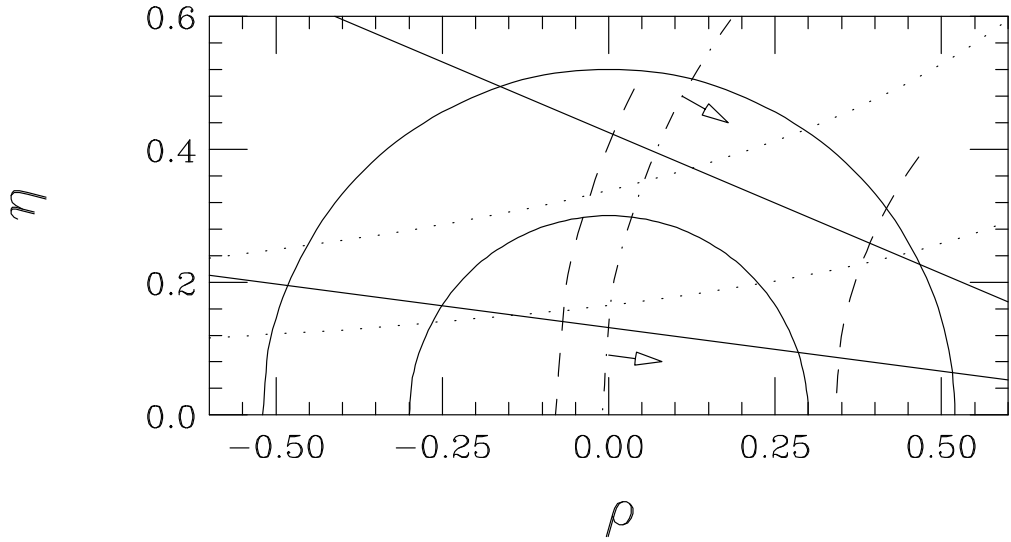}}
\caption{Region of $(\rho,\eta)$ specified by  $\pm 1 \sigma$ constraints on
CKM parameters.  Solid semicircles:  $|V_{ub}/V_{cb}|$.  Dotted hyperbolae:
$\epsilon_K$.  Dashed arcs:  $\Delta m_d$.  Dash-dotted arc: 95\% c.l. lower
limit on $\Delta m_s$.  Rays: $\sin 2 \beta \equiv \sin 2 \phi_1$.
\label{fig:re}}
\end{figure}

The constraints are plotted on the $(\rho,\eta)$ plane in Fig.\ \ref{fig:re}.
Also shown are the $\pm 1 \sigma$ bounds on $\sin 2 \beta \equiv \sin 2
\phi_1$ from an
average $0.49 \pm 0.23$ \cite{JRTASI} of OPAL, ALEPH, CDF, BaBar, and BELLE
values.  There is no contradiction (yet)!

We comment further on decay constants.  (1) The ratios $f_{B_s}/f_B$ and
$f_{D_s}/f_D$ should be very similar \cite{lat,Grin,CPPACS}.  The
value of $f_{D_s}$ has been measured to be $251 \pm 30$ MeV \cite{lat}.
The measurement of $f_D$ should be a first priority for a charm factory
producing $\psi(3770) \to D^+ D^-$, in which one of the charged $D$'s is
used for tagging and the other decays to $\mu \nu$. 
(2) The nonrelativistic quark model implies \cite{NRFM}
$|f_M|^2 = 12 |\Psi(0)|^2/M_M$ for the decay constant $f_M$ of
a meson $M$ of mass $M_M$ composed of a quark-antiquark pair with relative
wave function $\Psi(\vec{r})$.  One estimates the ratios of $|\Psi(0)|^2$
in $D$ and $D_s$ systems from strong hyperfine splittings.  Since $M(D^{*+})
- M(D^+) \simeq M(D_s^{*+}) - M(D_s^+)$, one expects $|\Psi(0)|_{Q \bar d}^2
/m_d \simeq |\Psi(0)|_{Q \bar s}^2/m_s$ for mesons containing a heavy quark
$Q$.  In constituent-quark models \cite{GasR} $m_d/m_s \simeq
0.64$, so $f_{Q \bar d}/f_{Q \bar s} \simeq \sqrt{0.64} = 0.8$.
(3) The measurement of $F_B$ depends on determination of
$B^+ \to \tau^+ \nu_\tau$ or $B^+ \to \mu^+ \nu_\mu$ to
sufficent accuracy \cite{JRCKM,fut}.  One predicts
\beq
{\cal B} \left\{ \begin{array}{c} {\cal B}(B^+ \to \mu^+ \nu_\mu) \\
         {\cal B}(B^+ \to \tau^+ \nu_\tau) \end{array} \right\} = 
 \left\{ \begin{array}{c} 2.5 \times 10^{-7} \\
 5.7 \times 10^{-5} \end{array} \right\} \left(
\frac{f_B}{200~{\rm MeV}} \right)^2 \left| \frac{V_{ub}}{0.003} \right|^2~~~.
\eeq
The ratio $\Gamma(B^+ \to \ell^+ \nu_\ell)/\Delta m_d$ is proportional to
$|V_{ub}/\sqrt{B_B} V_{td}|^2$; common factors of
$f_B$ cancel.  Given $B_B$, this provides $r \equiv |V_{ub}/V_{td}|$.
A 10\% measurement of $r$ would require (e.g.) 25 $B^+ \to \mu^+ \nu_\mu$
decays, or 100 million charged $B$'s (or $B \bar B$ pairs).

An estimate quoted in \cite{JRCKM,fut} concludes that in the year 2003
the allowed region in $(\rho,\eta)$ space may involve errors
of roughly $\pm 0.05$ on each parameter.  In contrast to some of the
more optimistic estimates \cite{FS}, I do not think we are there yet.

\section*{Acknowledgments}

I wish to thank Yoram Rozen, Peter Schlein, and the other organizers of Beauty
2000 for an enjoyable and informative conference, and
Prof.\ K. T. Mahanthappa at the University of Colorado
and colleagues at Cornell University and the Technion for gracious hospitality
and fruitful interactions.  This work was supported in part by the United
States Department of Energy through Grant No.\ DE FG02 90ER40560, and in part
by the U. S. -- Israel Binational Science Foundation through Grant No.\
98-00237.

% Journal and other miscellaneous abbreviations for references
\def \ajp#1#2#3{Am.\ J. Phys.\ {\bf#1} (#3) #2}
\def \apny#1#2#3{Ann.\ Phys.\ (N.Y.) {\bf#1} (#3) #2}
\def \app#1#2#3{Acta Phys.\ Polonica {\bf#1} (#3) #2}
\def \arnps#1#2#3{Ann.\ Rev.\ Nucl.\ Part.\ Sci.\ {\bf#1} (#3) #2}
\def \art{and references therein}
\def \cmts#1#2#3{Comments on Nucl.\ Part.\ Phys.\ {\bf#1} (#3) #2}
\def \cn{Collaboration}
\def \cp89{{\it CP Violation,} edited by C. Jarlskog (World Scientific,
Singapore, 1989)}
\def \econf#1#2#3{Electronic Conference Proceedings {\bf#1}, #2 (#3)}
\def \efi{Enrico Fermi Institute Report No.\ }
\def \epjc#1#2#3{Eur.\ Phys.\ J. C {\bf#1} (#3) #2}
\def \f79{{\it Proceedings of the 1979 International Symposium on Lepton and
Photon Interactions at High Energies,} Fermilab, August 23-29, 1979, ed. by
T. B. W. Kirk and H. D. I. Abarbanel (Fermi National Accelerator Laboratory,
Batavia, IL, 1979}
\def \hb87{{\it Proceeding of the 1987 International Symposium on Lepton and
Photon Interactions at High Energies,} Hamburg, 1987, ed. by W. Bartel
and R. R\"uckl (Nucl.\ Phys.\ B, Proc.\ Suppl., vol.\ 3) (North-Holland,
Amsterdam, 1988)}
\def \ib{{\it ibid.}~}
\def \ibj#1#2#3{~{\bf#1} (#3) #2}
\def \ichep72{{\it Proceedings of the XVI International Conference on High
Energy Physics}, Chicago and Batavia, Illinois, Sept. 6 -- 13, 1972,
edited by J. D. Jackson, A. Roberts, and R. Donaldson (Fermilab, Batavia,
IL, 1972)}
\def \ijmpa#1#2#3{Int.\ J.\ Mod.\ Phys.\ A {\bf#1} (#3) #2}
\def \ite{{\it et al.}}
\def \jhep#1#2#3{JHEP {\bf#1} (#3) #2}
\def \jpb#1#2#3{J.\ Phys.\ B {\bf#1} (#3) #2}
\def \kaon{{\it Kaon Physics}, edited by J. L. Rosner and B. Winstein,
University of Chicago Press, 2000.}
\def \lg{{\it Proceedings of the XIXth International Symposium on
Lepton and Photon Interactions,} Stanford, California, August 9--14 1999,
edited by J. Jaros and M. Peskin (World Scientific, Singapore, 2000)}
\def \lkl87{{\it Selected Topics in Electroweak Interactions} (Proceedings of
the Second Lake Louise Institute on New Frontiers in Particle Physics, 15 --
21 February, 1987), edited by J. M. Cameron \ite~(World Scientific, Singapore,
1987)}
\def \kdvs#1#2#3{{Kong.\ Danske Vid.\ Selsk., Matt-fys.\ Medd.} {\bf #1},
No.\ #2 (#3)}
\def \ky85{{\it Proceedings of the International Symposium on Lepton and
Photon Interactions at High Energy,} Kyoto, Aug.~19-24, 1985, edited by M.
Konuma and K. Takahashi (Kyoto Univ., Kyoto, 1985)}
\def \mpla#1#2#3{Mod.\ Phys.\ Lett.\ A {\bf#1} (#3) #2}
\def \nat#1#2#3{Nature {\bf#1} (#3) #2}
\def \nc#1#2#3{Nuovo Cim.\ {\bf#1} (#3) #2}
\def \nima#1#2#3{Nucl.\ Instr.\ Meth. A {\bf#1} (#3) #2}
\def \np#1#2#3{Nucl.\ Phys.\ {\bf#1} (#3) #2}
\def \npps#1#2#3{Nucl.\ Phys.\ Proc.\ Suppl.\ {\bf#1} (#3) #2}
\def \npbps#1#2#3{Nucl.\ Phys.\ B Proc.\ Suppl.\ {\bf#1} (#3) #2}
\def \os{XXX International Conference on High Energy Physics, Osaka, Japan,
July 27 -- August 2, 2000}
\def \PDG{Particle Data Group, D. E. Groom \ite, \epjc{15}{1}{2000}}
\def \pisma#1#2#3#4{Pis'ma Zh.\ Eksp.\ Teor.\ Fiz.\ {\bf#1} (#3) #2 [JETP
Lett.\ {\bf#1} (#3) #4]}
\def \pl#1#2#3{Phys.\ Lett.\ {\bf#1} (#3) #2}
\def \pla#1#2#3{Phys.\ Lett.\ A {\bf#1} (#3) #2}
\def \plb#1#2#3{Phys.\ Lett.\ B {\bf#1} (#3) #2}
\def \pr#1#2#3{Phys.\ Rev.\ {\bf#1} (#3) #2}
\def \prc#1#2#3{Phys.\ Rev.\ C {\bf#1} (#3) #2}
\def \prd#1#2#3{Phys.\ Rev.\ D {\bf#1} (#3) #2}
\def \prl#1#2#3{Phys.\ Rev.\ Lett.\ {\bf#1} (#3) #2}
\def \prp#1#2#3{Phys.\ Rep.\ {\bf#1} (#3) #2}
\def \ptp#1#2#3{Prog.\ Theor.\ Phys.\ {\bf#1} (#3) #2}
\def \rmp#1#2#3{Rev.\ Mod.\ Phys.\ {\bf#1} (#3) #2}
\def \rp#1{~~~~~\ldots\ldots{\rm rp~}{#1}~~~~~}
\def \si90{25th International Conference on High Energy Physics, Singapore,
Aug. 2-8, 1990}
\def \slc87{{\it Proceedings of the Salt Lake City Meeting} (Division of
Particles and Fields, American Physical Society, Salt Lake City, Utah, 1987),
ed. by C. DeTar and J. S. Ball (World Scientific, Singapore, 1987)}
\def \slac89{{\it Proceedings of the XIVth International Symposium on
Lepton and Photon Interactions,} Stanford, California, 1989, edited by M.
Riordan (World Scientific, Singapore, 1990)}
\def \smass82{{\it Proceedings of the 1982 DPF Summer Study on Elementary
Particle Physics and Future Facilities}, Snowmass, Colorado, edited by R.
Donaldson, R. Gustafson, and F. Paige (World Scientific, Singapore, 1982)}
\def \smass90{{\it Research Directions for the Decade} (Proceedings of the
1990 Summer Study on High Energy Physics, June 25--July 13, Snowmass,
Colorado),
edited by E. L. Berger (World Scientific, Singapore, 1992)}
\def \tasi{{\it Testing the Standard Model} (Proceedings of the 1990
Theoretical Advanced Study Institute in Elementary Particle Physics, Boulder,
Colorado, 3--27 June, 1990), edited by M. Cveti\v{c} and P. Langacker
(World Scientific, Singapore, 1991)}
\def \yaf#1#2#3#4{Yad.\ Fiz.\ {\bf#1} (#3) #2 [Sov.\ J.\ Nucl.\ Phys.\
{\bf #1} (#3) #4]}
\def \zhetf#1#2#3#4#5#6{Zh.\ Eksp.\ Teor.\ Fiz.\ {\bf #1} (#3) #2 [Sov.\
Phys.\ - JETP {\bf #4} (#6) #5]}
\def \zpc#1#2#3{Zeit.\ Phys.\ C {\bf#1} (#3) #2}
\def \zpd#1#2#3{Zeit.\ Phys.\ D {\bf#1} (#3) #2}

\newpage

\end{document}